\documentclass[aps,prd,reprint,superscriptaddress,nofootinbib,longbibliography]{revtex4-1}

\usepackage{amsmath,amssymb, amsthm,amstext}
\usepackage{natbib}
\usepackage{graphicx}
\usepackage{color}
\usepackage{array, enumerate}
\usepackage{bm}
\usepackage{multirow}
\usepackage[breaklinks,colorlinks,citecolor=blue]{hyperref}
\usepackage{braket}
\usepackage{txfonts}
\usepackage{physics}
\usepackage[normalem]{ulem}

\def\be{\begin{equation}}
\def\ee{\end{equation}}
\newcommand{\comment}[1]{}

\begin{document}

\title{Extreme Mass Ratio Inspirals in Light of Quasi-periodic Eruptions:\\ Milli-Hertz Gravitational Wave Background}

\author{Brennen Black}
\email{brennenblack@sjtu.edu.cn}
\affiliation{Tsung-Dao Lee Institute, Shanghai Jiao-Tong University, Shanghai, 520 Shengrong Road, 201210, China}
\affiliation{School of Physics \& Astronomy, Shanghai Jiao-Tong University, Shanghai, 800 Dongchuan Road, 200240, China}
\author{Xian Chen}
\email{xian.chen@pku.edu.cn}
\affiliation{Department of Astronomy, School of Physics, Peking University, Beijing 100871, P. R. China}
\affiliation{Kavli Institute for Astronomy and Astrophysics at Peking University, Beijing 100871, P. R. China}

\author{Zhen Pan}
\email{zhpan@sjtu.edu.cn}
\affiliation{Tsung-Dao Lee Institute, Shanghai Jiao-Tong University, Shanghai, 520 Shengrong Road, 201210, China}
\affiliation{School of Physics \& Astronomy, Shanghai Jiao-Tong University, Shanghai, 800 Dongchuan Road, 200240, China}
\begin{abstract}
Quasi-periodic eruptions (QPEs) are repeated X-ray bursts originating in galactic nuclei. Of the many proposed models, the favored model is the disk-collision model in which a stellar mass orbiter collides with a disk formed from a tidal disruption event, generating flares twice per orbit. In this model QPEs are tracers of circular extreme mass ratio inspirals (EMRIs) and can be used to infer the EMRI formation rate and estimate their contribution to the stochastic gravitational wave background (SGWB) in the Laser Interferometer Space Antenna (LISA) band. Whether the secondary is a stellar-mass black hole or a main sequence star is still debated and leads to different results for the EMRI rate and SGWB. We obtain fiducial rates --- subject to systematic uncertainties --- of $R_{\rm SE} = 2.88\times10^{-6}$ per galaxy per year for stellar EMRIs and $R_{\rm BHE} = 6.07\times10^{-6}$ per galaxy per year for black hole EMRIs, then  estimate their contribution to the SGWB. We find that only black hole EMRIs contribute to the 1 - 10 milliHertz band resolvable by LISA, and depending on the secondary mass and formation radius can contribute from just below the LISA sensitivity curve to roughly two orders of magnitude above it. Stellar EMRIs, being tidally disrupted before reaching the 1 - 10 milliHertz band, only contribute to sub-milliHertz frequencies and remain below the LISA sensitivity curve.
\end{abstract}
\date{\today}

\maketitle

\section{\label{sec:level1}Introduction}
Quasi-periodic eruptions (QPEs) are short, repeating bursts of X-ray emissions originating from galactic nuclei. These transients, with luminosities between $10^{40}-10^{43}$ ergs s$^{-1}$ and temperature $kT \sim 100 - 250$ eV, produce recurring flares anywhere from a few hours to tens of days apart, and over a lifespan of roughly a few years \cite{Sun2013,Arcodia2021,Arcodia2022,Arcodia2024,Chakraborty2021,Evans2023,Giustini2020,Guolo2024,HernndezGarca2025,Miniutti2019,Miniutti2023,Miniutti2023-2,Nicholl2024,Quintin2023,Wevers2022,Baldini2026}. Analysis of the confirmed sources suggest that QPEs typically form in galaxies with a central supermassive black hole (SMBH) in the lower mass end with $\log_{10}(M_\bullet/M_{\odot}) \sim 5-7$ \cite{Miniutti2019,Giustini2020,Wevers2022,Arcodia2021,Arcodia2022,HernndezGarca2025}. 
Despite apparent diversities in QPEs, a number of unifying features are becoming increasingly central to the QPE picture. Many recent studies have drawn connections between the morphological properties of QPEs and tidal disruption events (TDEs), with some QPE sources having confirmed TDE precursors \cite{Wu2025,Nicholl2024,Chakraborty2025,Quintin2023,Chakraborty2021}. In particular, QPEs overwhelmingly belong to the same population as TDE hosts -- low mass black holes in post-starburst or quiescent galaxies. With more in-depth studies of these host galaxies, it was discovered that many contain an extended emission-line region and/or TDE infrared echo, indicating a recently faded AGN \cite{Wevers2024,Xiong2025,Jiang2025,Wu:2025vgt,Baldini2026}. These studies further confirm the connection between QPEs, TDEs and recent AGN activities.

\indent There are predominantly three QPE models which have been explored extensively in the literature. The first, used primarily to explain GSN 069, is the partial tidal stripping of an eccentric ($e \gtrsim 0.9$) white dwarf or main-sequence star \cite{King2020,King2022,Wang2022}. The second model explores how radiation pressure instabilities of the inner regions of a SMBH's accretion disk generates repeating flares \cite{Pan2022,Sniegowska2023}. Finally, the third model considers disk collisions as the source of the observed soft X-ray bursts \cite{Linial2023,Franchini2023,Xian2021}. The model of periodic electromagnetic bursts due to star-disk collisions first appeared in 2010 \cite{Dai2010}, and consists of a TDE that generates a disk of accreting matter through which a compact secondary on a tight, low to mildly eccentric orbit ($e \lesssim 0.3$) crosses \cite{Zhou2024,Zhou2024b,Zhou2024b,Xian2021,Franchini2023}, generating X-ray emissions twice per orbit at quasi-periodic intervals.

The disk collision model is by no means complete, but it has quickly become the favored model. The most obvious point of favor is the proposition of a source for the long-short timing patterns $T_{\rm long}, T_{\rm short}$ between successive QPE flares in e.g., GSN 069 \cite{Miniutti2019}. Zhou et al. \cite{Zhou2024} further pointed out that both  $T_{\rm long}$ and $ T_{\rm short}$ are varying with time
while their sum $T_{\rm long}+ T_{\rm short}$ of consecutive flares remains approximately a constant. This discovery strongly indicates that $T_{\rm long}+ T_{\rm short}$ is the fundamental period of QPEs, and there are two eruptions produced per period, which is consistent 
with simulations of star/BH disk collisions  \cite{Liu2026,Huang2026}. 
A slightly eccentric orbit which crosses a TDE disk twice per orbit naturally provides this long-short pattern without the requirement of additional physics. Timing analyses of confirmed QPE sources indeed support low to mild eccentricities \cite{Xian2021,Zhou2024,Zhou2024b,Zhou:2024vwj,Franchini2023,Chakraborty2024}.


The disk collision model makes QPEs a valuable tool for probing astrophysical processes around SMBHs. The orbit of the secondary directly conveys information about the local spacetime around the central SMBH. Several works have focused on using QPE timing data to reconstruct the secondary's orbit and constrain the central SMBH's mass and spin \cite{Xian2021, Zhou2024,Zhou2024b,Zhou:2024vwj, Zhou2025, Chakraborty2025b}. Other applications include using QPEs as a way to study TDE disks around the central SMBH \cite{Suzuguchi2026,Mondek2026,Guolo2025,Guolo2025b}. Of relevance to this work is that QPEs in the disk collision model are tracers of extreme mass ratio inspirals (EMRIs) and could be used to infer the EMRI formation channels \cite{Zhou2024,Zhou2024b} and the EMRI formation rate \cite{Arcodia2024b, Allievi2026}. The evolution of these particular EMRIs are dominated by the radiation of milliHertz (mHz) frequency gravitational waves, falling into the sensitive band of the space-born gravitational wave missions Laser Interferometer Space Antenna (LISA) \cite{AmaroSeoane2017}, Tianqin \cite{Mei2020,Lu2019,Hu2018,Luo2016}, and Taiji \cite{Hu2017,Luo2020}. The low frequency emissions of these systems indicate they will not be individually resolvable by these instruments but instead will contribute to the stochastic gravitational wave background (SGWB). \cite{Chen2022} studies how eccentric white dwarf QPEs contribute to the SGWB, but a similar study assuming the disk collision model has yet to be done. Such studies make clear that beyond serving as observational confirmation of current theories of EMRI formation, QPEs hold the unique power to provide the first observationally-motivated model of the SGWB.\\
\indent A critical discussion of the disk collision model is the nature of the secondary, specifically whether the secondary is a stellar-mass black holes (sBHs) or a main-sequence star. \cite{Linial2023} provide an analytical framework for QPE formation and evolution assuming the assumption that the secondary is a main-sequence star, and in subsequent years others used radiation-hydrodynamic simulations to study the spectral properties of QPEs, finding some success in explaining observed QPEs \cite{Huang2025,Huang2026,Jankovic2026,Vurm2025}. In particular, the work of \cite{Linial2023} and various numerical studies (for example, \cite{Dodd2025,Lam2025}) find that sBHs on roughly circular, perpendicular orbits to the TDE disk do not have a sufficiently large Bondi radius and therefore cannot replicate the observed peak luminosities of confirmed sources unless they are intermediate mass black holes. Other studies point out that sBHs on prograde orbits and with inclinations less than roughly $15^{\circ}-20^{\circ}$ have lower relative velocities and therefore obtain a sufficiently large Bondi radius \cite{Franchini2023,Liu2026}.\\
\indent The criticality of the distinction between sBHs and stars as the secondary, in this context of the present work, lies in their respective lifetimes. A sBH can survive multiple TDE disks and produce multiple QPEs on the coalescence timescale. In contrast, stars lose a significant fraction of it's mass via mass ablation on timescales of decades \cite{Linial2023}, and therefore are unlikely to persist over many TDE disks. The difference between these two timescales can reach up to 6 orders of magnitude, thereby warranting further investigation into the effects on the EMRI formation rate and SGWB.\\
\indent In this paper, we will use the disk collision model of QPE formation along with observationally-motivated QPE data to calculate both the EMRI formation rate and SGWB. The paper will be organized as follows. In Section \ref{sec:level2} we calculate the EMRI formation rate using the disk collision model. We will calculate the EMRI formation rate assuming that all QPEs are black hole EMRIs (BHEs), and then again assuming all QPEs are stellar EMRIs (SEs). In Section \ref{sec:level3} we use the results of Section \ref{sec:level2} to run simulations to infer the SGWB from a mock population of EMRIs. We explore how the mass distribution of the secondary affects the SGWB, as well as the radius of the TDE disk. In Section \ref{sec:level4} we will summarize the results and discuss avenues of future research.
\section{\label{sec:level2}EMRI Formation Rates}
The disk collision model can be decomposed into three events that are roughly independent processes: EMRI formation, TDE disk formation, and QPE formation and detection. First, one must have a low-eccentricity EMRI in the galactic nucleus. Once a suitable EMRI system is formed it must coincide with a TDE disk. Initially the TDE disk forms at around twice the tidal radius \cite{Linial2023,Cannizzo1990,Rees1988,Ulmer1999,Strubbe2009}. The disk is viscous and begins to expand; some mass migrates towards the innermost stable circular orbit (ISCO), and by conservation of angular momentum mass also expands outwards \cite{LyndenBell1974,Cannizzo1990}. In general not every disk will intersect with the EMRI orbit. The coexistence of an EMRI and a TDE disk does not imply that system will generate QPEs. In the systems where the disk expands and eventually crosses the secondary's orbit, we must then detect and confirm the resulting emissions as a QPE. The probability of detecting and confirming a QPE source comes down to the algorithm used to search available light-curve data from X-ray telescopes. An example of one such algorithm is given in \cite{Arcodia2024b}. \cite{Chakraborty2025} builds on this work in their Appendix B by exploring how minimum flare luminosity, observation time, and number of observed peaks impact the detection probability.\\
\indent Another important result of \cite{Chakraborty2025} is the use of QPEs with confirmed optical TDE precursors to determine the number of TDE host sites that are expected to generate observable QPEs within 5 years of the TDE. This study provides a strong, observationally motivated constraint on the ratio of the QPE rate to the TDE rate, namely $q \equiv R_{\text{QPE}}/R_{\text{TDE}} \approx 0.09$, which is related to the EMRI formation rate. Such a relationship however depends critically on whether the secondary is a black hole or a main-sequence star.\\
\indent Arcodia et. al. 2024 \cite{Arcodia2024b} use QPEs discovered by the eROSITA X-ray instrument to estimate the volumetric QPE rate, and further use those results to infer the EMRI formation rate. Taking the lifetime of the QPE to be roughly 10 years, they determine a QPE rate of roughly $0.4\times10^{-5}$ events per galaxy per year. This study provides another observationally motivated estimate of the QPE rate, and claims the QPE rate serves as a strict lower bound to the EMRI rate. While this assumption is justifiable for stellar secondaries, it does not apply to sBHs, which are capable of surviving more than one TDE disk. Using the result of \cite{Chakraborty2025}, we look to compare the resultant EMRI rate to that of \cite{Arcodia2024b}.\\
\indent To first order, and under the assumption that the disk collision model is unambiguously favored over other models, the QPE rate $R_{\rm QPE}$ is related to the EMRI rate $R_{\rm EMRI}$ by
\begin{equation}\label{eq:GeneralRQPEFormula}
    R_{\rm QPE} = \lambda R_{\rm EMRI}N_{\rm disk},
\end{equation}
where $N_{\rm disk}$ is the expected number of disks encountered by a given system and $\lambda$ is a geometric constraint that applies to BHE-QPEs but not SE-QPEs. Eq.\eqref{eq:GeneralRQPEFormula} applies to both BHE-QPEs and SE-QPEs, but the expression for $N_{\rm disk}$ reflects the bottleneck to QPE production and is thus different for each type of system. We start by assuming that all QPEs are SEs. Under this assumption, there are no orbital constraints needed for the QPE to be detectable \cite{Linial2023} so we take $\lambda = 1$. Furthermore, stellar secondaries are subjected to mass ablation at each disk crossing. These stars are estimated to be destroyed after roughly 10 years which is roughly equal to the expected lifetime of the TDE disk \cite{Linial2023}. Stellar secondaries are light enough such that each one is nearly guaranteed to coincide with a TDE disk, and thus the bottleneck for QPE production is whether or not the stellar secondary can survive its first disk. For these systems, given the current estimates provided earlier, it is justifiable that $N_{\rm disk} \approx 1$, and therefore the SE formation rate $R_{\rm SE}$ is given by the QPE rate. Equivalently,
\begin{align}\label{eq:RSEFormula}
    R_{\rm SE} &= R_{\rm TDE}\times\left(\frac{R_{\rm QPE}}{R_{\rm TDE}}\right)= 2.88\times10^{-6}\text{ gal}^{-1}\text{ yr}^{-1} \nonumber\\
    &\times \left(\frac{R_{\rm QPE}/R_{\rm TDE}}{0.09}\right)\left(\frac{R_{\rm TDE}}{3.2\times10^{-5}\text{ gal}^{-1}\text{ yr}^{-1}}\right),
\end{align}
where we've used Yao et. al. 2023's \cite{Yao2023} estimate of the TDE formation rate $R_{\rm TDE}$ in order to be consistent with \cite{Chakraborty2025}. These results are consistent with the assumption and lower bound provided in \cite{Arcodia2024b}.\\
\indent Now we assume that all QPEs are BHEs. The setup changes in a couple of important ways. First is the necessity for the orbital constraint $\lambda$. \cite{Linial2023} argues that black hole secondaries in orbits perpendicular to the TDE disk move too fast relative to the disk, and therefore have a Bondi radius that is insufficient to reproduce the peak luminosities observes via X-ray telescopes. \cite{Franchini2023} and \cite{Allievi2026} note that if the secondary has a prograde orbit relative to the disk and has an inclination no larger than roughly $20^{\circ}$, the Bondi radius is sufficiently large as to reproduce the peak luminosities. Following these analyses, we assume an isotropic distribution of inclinations and require an inclination of no larger than $20^{\circ}$ which gives $\lambda \approx 0.06$. Another important change is that sBHs are generally heavier than stellar secondaries, and thus coalesce faster. A significant portion of BHEs may not experience a TDE disk before merging, and this becomes the more significant bottleneck to QPE production.\\
\indent The role of the coalescence time for BHEs is more subtle than that for SEs, and its calculation thereby deserves its own discussion. The coalescence time of an EMRI system as a function of semi-major axis is given by \cite{Peters1964}
\begin{align}\label{eq:sec2tCoalFormula}
    t_{\text{coal}}(a) &= \frac{5c^5}{256G^3}\frac{a^4}{M_\bullet^2M_{s}}\ ,
\end{align}
when the EMRI's orbital evolution is dominated by the emission of gravitational waves. 
We start by adopting $M_\bullet = 10^{6}M_{\odot}$ as a standard primary mass given that the SMBH density is log-uniform with respect to $M_{\bullet}$ \cite{Lyu2026}. The distribution of the secondary mass is unknown, but \cite{Franchini2023} notes that the black hole mass must not greatly exceed 100 $M_{\odot}$; we therefore take the secondary mass $M_{s}$ to be $M_{s} = 100M_{\odot}$ as our fiducial number, but note that realistically the secondary mass can be as low as $10 M_{\odot}$. Since the distribution is unknown we simply choose one extreme as a fiducial number, though we note that we consider both extremes in later analysis. In the literature, QPE timing analyses \cite{Franchini2023,Zhou:2024vwj,Zhou2024,Zhou2024b,Giustini2024b} estimate that the semi-major axes range from 50$R_{\rm g}$ up to roughly 1000$R_{\rm g}$, with a majority of estimates falling at or above a few hundred $R_{\rm g}$, where $R_g = GM_\bullet/c^2$. Physically, there are constraints on the orbital size of QPE EMRIs, particularly that $a \lesssim r_{\rm disk}$, where $r_{\rm disk}$ is the TDE disk radius.

One can picture that a TDE disk forms initially at a radius $r_{\rm disk,0}$ given by \cite{Linial2023,Cannizzo1990,Rees1988,Ulmer1999,Strubbe2009}
\begin{equation}\label{eq:initialRdisk}
    r_{\text{disk},0} \approx 2R_{\odot}\left(\frac{M_\bullet}{M_{\odot}}\right)^{1/3},
\end{equation}
where one assumes the progenitor is preferentially a solar-type star \cite{Mockler2022}. Over the lifetime of the disk, some matter falls into the SMBH and so by conservation of angular momentum the outer edge of the disk expands \cite{LyndenBell1974,Cannizzo1990}. Since the range of primary masses is fairly well understood, we can explicitly calculate the range of $r_{\rm disk,0}$ in units of $R_g$. For a primary of mass $M_{\bullet} \in [10^{5},10^{7}]M_{\odot}$, using Eq.\eqref{eq:initialRdisk} one can calculate $r_{\rm disk,0} \in [50, 500]R_{\rm g}$, with $r_{\rm disk,0}\sim100R_{\rm g}$ when $M_{\bullet} = 10^{6}M_{\odot}$. The expansion of the disk may further enlarge the disk size to a few times $10^3 R_{\rm g}$.

Starting with a low initial size $r_{\rm disk, 0}$, the TDE disk gradually expands to the EMRI orbit with $r_{\rm disk}\approx a$, when QPEs are produced. We can therefore write the reference coalescence time to be
\begin{align}\label{eq:sec2tCoal}
    t_{\text{coal}}(r_{\rm disk}) = 2.47\times 10^5\ {\rm yr}\times \left(\frac{M_\bullet}{10^6 M_\odot}\right)^{2} \left(\frac{M_s}{100 M_\odot}\right)^{-1} \left(\frac{r_{\rm disk}}{300 R_{\rm g}}\right)^{4}.
\end{align}
Furthermore, we note that both lowering the secondary mass and increasing the TDE disk radius serve to increase the reference coalescence time.

Returning to the calculation of the BHE formation rate $R_{\rm BHE}$, we can write the expected number of disks the BHE will experience as the product of the lifetime of the secondary and the per-year formation rate of TDE disks
\begin{equation}\label{eq:BHENdisk}
    N_{\rm disk} = R_{\rm TDE}t_{\rm coal}(r_{\rm disk}).
\end{equation}
For the choice of fiducial parameters discussed in the previous paragraphs and the TDE rate of \cite{Yao2023} this value is roughly $N_{\rm disk} \sim 8$; larger than unity, though not by much. However, the existing parameter space allows for numbers much larger than unity. Plugging Eq.~\eqref{eq:BHENdisk} into Eq.~\eqref{eq:GeneralRQPEFormula}, and making sure to keep the orbital constraint $\lambda$ since it is no longer unity, we can easily solve for $R_{\rm BHE}$
\begin{align}\label{eq:R_BHE}
    R_{\rm BHE} &= \frac{R_{\rm QPE}/R_{\rm TDE}}{\lambda t_{\rm coal}(r_{\rm disk})} = 6\times10^{-6}\text{ gal}^{-1}\text{ yr}^{-1}\ \nonumber \\
    &\times \left(\frac{R_{\rm QPE}/R_{\rm TDE}}{0.09}\right)
    \left(\frac{\lambda}{0.06}\right)^{-1}\left(\frac{t_{\rm coal}(r_{\rm disk})}{2.47\times10^{5}\text{ yr}}\right)^{-1},
\end{align}
where we've again used the fiducial ratio $R_{\rm QPE}/R_{\rm TDE} \sim 0.09$ \cite{Chakraborty2025}.

This fiducial value is crucially not a lower bound, since we have taken the QPE production criterion into account. According to Eq.~(\ref{eq:R_BHE}), a number of updated choices on the free parameters can drive this rate down (for example, a larger choice of $r_{\rm disk}$ or a less extreme choice of the secondary mass) by many orders of magnitude, showing that the EMRI rate may not be so large. It is especially important to provide a comparison of our predicted rates ($\sim10^{-6}$ gal${}^{-1}$ yr${}^{-1}$) to specific formation channels. A number of works have been published exploring formation rates for channels such as two-body relaxation, AGN disk migration and Hills' mechanism, both on the per MBH per year scale and via population averages. In particular, \cite{Mancieri2025,BarOr2016, Broggi2022} show that the two-body relaxation EMRI rate is of order $10^{-9}-10^{-6}$ gal${}^{-1}$ yr${}^{-1}$ depending on the details of the model, \cite{Pan2021a,Pan2021b} deduce that AGN disk migration (the wet channel) yields an EMRI rate of up to order $10^{-6}$ gal${}^{-1}$ yr${}^{-1}$, while \cite{Metzger2022,Linial2023b} infers the Hills' mechanism EMRI rate to be of order $10^{-7}-10^{-5}$ gal${}^{-1}$ yr${}^{-1}$ for Milky Way-like galaxies. The EMRI rates presented in Eq.~\eqref{eq:RSEFormula} and Eq.~\eqref{eq:R_BHE} do not assume \textit{a priori} a specific formation channel, but are nonetheless consistent with the literature modulo some specific choices of conventions.
\section{\label{sec:level3}Stochastic Gravitational Wave Background}
The results of Section \ref{sec:level2} provide a new, data-driven insight into the current population of EMRIs, and simultaneously opens the door to study properties of these systems on a global scale. Of great interest to those studying gravitational waves is the SGWB. While LISA will be able to clearly resolve significantly strong merging EMRI systems in the 1 mHz - 10 mHz frequency range \cite{Bonetti2020}, there are a number of systems (QPEs included) that will not be individually resolvable by LISA. The collection of these systems generate a continuous, noisy background that can distort individual signals, and thus require careful study. While our current understanding of QPE formation and evolution is not strong enough to provide definitive results, we can obtain a fair estimate by simulating a population of EMRIs using the Monte-Carlo technique whose size is motivated by the results of Section \ref{sec:level2}. For the remainder of this section, we assume that each system has zero eccentricity.\\
\indent The quantity of interest is the characteristic strain of the gravitational wave background as a function of frequency. Following \cite{Bonetti2020}, we use  
\begin{equation}\label{eq:SGWBStrain}
    h_{\text{c,gwb}}^{2}(f) = \int dzd\mathcal{M}\frac{d^3N}{dzd\mathcal{M}d\ln(f_{\text{orb}})}\min\left(\frac{t_{\rm res}(f)}{t_{\rm LISA}}, 1\right)h^{2}(f),
\end{equation} 
where
\begin{equation}\label{eq:SingleSystemStrain}
    h^2(f) = \frac{32(G\mathcal{M}(1+z))^{10/3}f^{4/3}}{5c^{8}d_{L}^2},
\end{equation}
and
\begin{equation}\label{eq:perFreqResTime}
    t_{\rm res}(f) \equiv \frac{f}{\dot{f}} = \frac{5}{96}\left(\frac{c^3}{G\mathcal{M}(1+z)}\right)^{5/3}(\pi f)^{-8/3}.
\end{equation}
In the above three equations, $z$ is the redshift, $\mathcal{M}$ is the chirp mass, $f$ is the observed GW frequency, $f_{\text{orb}}$ the orbital frequency of a given EMRI system (related to $f$ by $f(1+z) = 2f_{\text{orb}}$), $t_{\text{LISA}} = 4$ yr the duration of a LISA mission, and $d_{L}$ is the luminosity distance to the source. Eq.\eqref{eq:SGWBStrain} can be interpreted as the expectation of the observable
\begin{equation}\label{eq:MCObservable}
    G(z,\mathcal{M},\ln(f_{\text{orb}});f) \equiv \min\left(\frac{t_{\rm res}(f)}{t_{\rm LISA}}, 1\right)h^{2}(f)
\end{equation}
over the joint distribution
\begin{equation}\label{eq:MCDistribution}
    \rho(z,\mathcal{M},\ln(f_{\text{orb}})) = \frac{d^3N}{dzd\mathcal{M}d\ln(f_{\text{orb}})},
\end{equation}
If we sample $N$ systems from the distribution $\rho(z,\mathcal{M},\ln(f_{\text{orb}}))$ then $h_{\text{c,gwb}}^{2}(f)$ is well-approximated by the Monte-Carlo estimator
\begin{equation}\label{eq:MCEstimator}
    \hat{h}^{2}(f) = \frac{N_{\text{total}}}{N}\sum_{\mathcal{I}}G(z_{i},\mathcal{M}_{i},\ln(f_{\text{orb,i}});f),
\end{equation}
where $N_{\rm total}$ is the normalization factor to Eq.~\eqref{eq:MCDistribution}, and $\mathcal{I}$ is the set
\begin{equation}\label{eq:summationSet}
    \mathcal{I} = \{\phantom{i}i\phantom{i}|f(1+z_{i})/2\in[f_{\rm orb,min,i}, f_{\rm orb,max,i}]\}
\end{equation}
The necessity of $\mathcal{I}$ lies in the fact that each system sweeps through a range of frequencies over the LISA mission duration. The minimum in Eq.~\eqref{eq:MCObservable} weights each system in an $f$-bin by the fraction of the time spent within a LISA mission in that particular $f$-bin, operating under the assumption that the system even exists in that $f$-bin. However, not every system will reach each $f$-bin; for example, not every system will reach the bins at $f \geq 10$ mHz. Instead, each system is initialized at an orbital frequency $f_{\rm orb,min} = f_{\rm orb}(t=0)$ (the details of this initialization are discussed below), and will reach a maximum orbital frequency given by $f_{\rm orb,max} = \min(f_{\rm orb,end}, f_{\rm orb}(t=t_{\rm LISA}))$. The term $f_{\rm orb,end}$ represents the final orbital frequency of the system before the EMRI is interrupted. For BHE-QPEs, $f_{\rm orb,end}$ is taken to be the orbital frequency at the ISCO, while for SE-QPEs $f_{\rm orb,end}$ is taken to be the orbital frequency at the tidal radius. A system will therefore exist in a given $f$-bin so long as $f(1+z_i)/2\in[f_{\rm orb,min,i},f_{\rm orb,max,i}]$.\\
\indent The joint distribution $\rho(z,\mathcal{M},\ln(f_{\text{orb}}))$ and it's normalization constant $N_{\text{total}}$ are directly related to the EMRI formation rates calculated in Section \ref{sec:level2}. The joint distribution can be separated as
\begin{equation}\label{eq:intermediateJointPDF}
    \rho(z,\mathcal{M},\ln(f_{\text{orb}})) = \frac{d^3N}{dzd\mathcal{M}d\ln(f_{\text{orb}})} = \frac{dV_{c}}{dz}\frac{f_{\text{orb}}}{\dot{f}_{\text{orb}}}\frac{d^3N}{dV_{c}d\tau d\mathcal{M}},
\end{equation}
where $V_{c}$ is the comoving volume at redshift $z$ and $\tau$ is time in the rest-frame of the SMBH. The volumetric EMRI rate $\mathcal{R}$ is defined as $\mathcal{R} = d^2N/dV_{c}d\tau$ and thus
\begin{equation}\label{eq:finalJointPDF}
    \rho(z,\mathcal{M},\ln(f_{\text{orb}})) = \frac{dV_{c}}{dz}\frac{f_{\text{orb}}}{\dot{f}_{\text{orb}}}\frac{d\mathcal{R}}{d\mathcal{M}}.
\end{equation}
From Eq.~\eqref{eq:finalJointPDF} it is clear that the joint PDF is the product of three uncorrelated probability densities corresponding to the three variables $z$, $\mathcal{M}$ and $\ln(f_{\text{orb}})$. Therefore, when initializing each system in the Monte-Carlo simulation, we can sample independently from each distribution to obtain the parameters of each system. Marginalization over all three probability densities will yield the normalization factor $N_{\text{total}}$, which is the total number of mock systems for a given volume in the $(z,\mathcal{M},\ln(f_{\text{orb}}))$ phase space. For a simulation up to some $z = z_{\text{max}}$, $N_{\rm total}$ is explicitly given by
\begin{equation}\label{eq:TotalSimulations}
    N_{\text{total}} = V_{c}(z_{\text{max}})\mathcal{R}\tau_{\text{res}}^{\rm fid},
\end{equation}
where $\tau^{\rm fid}_{\text{res}}$ is the residence time of a typical EMRI system. For arbitrary initial and final orbital frequencies $f_{\rm orb,min}$ and $f_{\rm orb,max}$ respectively, the residence time is given by
\begin{align}\label{eq:ResidenceTime}
    \tau_{\text{res}} = &-\frac{3}{8K}\left(f_{\text{orb,max}}^{-8/3} - f_{\text{orb,min}}^{-8/3}\right),\nonumber\\
    &\text{ where } K = \frac{96(2\pi)^{8/3}}{5}\left(\frac{G\mathcal{M}}{c^3}\right)^{5/3}.
\end{align}
The ‘fid’ superscript in $\tau_{\rm res}^{\rm fid}$ indicates that we use the fiducial parameters discussed in Section \ref{sec:level2} to calculate $f_{\rm orb,min}$ and $f_{\rm orb,max}$. That is, to determine $N_{\rm total}$ for a given choice of fiducial parameters, we calculate the Keplerian orbital frequency with those fiducial parameters for use in place of $f_{\rm orb,min}$, making sure to use the appropriate parameters for BHE-QPEs and SE-QPEs. For $f_{\rm orb,max}$ we use $f_{\rm orb,end}$ from above, again making sure to use the appropriate fiducial parameters for BHE-QPEs and SE-QPEs. The volumetric rate $\mathcal{R}$ can be straightforwardly related to the galactic EMRI rate $R$. Using the SMBH log-mass density of \cite{Lyu2026} (which is further supported by \cite{Yao2023,Gallo2019}), 
\begin{equation}\label{eq:SMBHlogMassDensity}
    \frac{dn}{d\log(M_\bullet)} = 0.005\text{ Mpc}^{-3}
\end{equation}
which is a good approximation for SMBHs within $10^{5}M_{\odot}$ and $10^{7}M_{\odot}$, one can integrate Eq.\eqref{eq:SMBHlogMassDensity} over the specified primary log-mass range to determine the number of SMBHs per volume. Multiplying this by the galactic EMRI rate (under the assumption that each galaxy holds only one SMBH) one obtains the volumetric EMRI rate $\mathcal{R} = 10^{-2}R$, where $R$ can be either the BHE or SE formation rate from Section \ref{sec:level2}. We further assume the standard $\Lambda$CDM cosmology to calculate the comoving volume (as well as for other cosmological quantities such as the luminosity distance $d_{L}$).\\
\indent We run two sets of Monte-Carlo simulations, one that assumes all QPEs are BHEs and another that assumes all QPEs are SEs (in accordance with Section \ref{sec:level2}). For both sets of simulations, we simulate up to $z_{\text{max}} = 20$ but note that convergence occurs at much lower redshifts. The distribution of the secondary mass is unknown, so we consider two separate scenarios for both BHE systems and SE systems. For BHEs, we perform one set of simulations where every secondary is 100 $M_{\odot}$ and another where every secondary is 10 $M_{\odot}$. For SEs we follow a similar protocol but with appropriately reduced masses: a maximum mass of 10 $M_{\odot}$ and a minimum mass of 1 $M_{\odot}$. We calculate $N_{\rm total}$ separately for each of the four types of simulations (BHE with $M_{s} = 10M_{\odot},100M_{\odot}$ and SE with $M_{s} = 1M_{\odot},10M_{\odot}$), which necessarily requires the recalculation of quantities such as $R_{\rm BHE}$ for BHE-QPEs and $\tau_{\rm res}^{\rm fid}$ for both BHE-QPEs and SE-QPEs. Due to computational restrictions we simulate $N = \min(10^{6}, N_{\text{total}})$, noting this is still consistent with Eq.\eqref{eq:MCEstimator}. For each individual system we first sample the primary mass from a log-uniform distribution between 5 and 7. To sample $z$ and $\ln(f_{\text{orb}})$ we use the method of inverse CDF sampling, motivated by the separability of the distributions in Eq.\eqref{eq:finalJointPDF}. In sampling $\ln(f_{\text{orb}})$ for each of the $N$ systems, we calculate the minimum and maximum orbital frequencies based on that system's sampled parameters and the requirements associated to assuming the secondary is a BHE or an SE. This means quantities such as the ISCO radius, tidal radius, and TDE disk radius are calculated separately for each individual system. To be clear, we reiterate that fiducial parameters are only used to calculate $N_{\rm total}$, whereas each individual system should calculate system-specific quantities based on the sampled parameters ($z,\mathcal{M},\ln(f_{\rm orb})$). The observed gravitational wave frequency $f_{i}$ can then be calculated using $f_{i} = 2f_{\text{orb,i}}/(1+z_{i})$. We then calculate Eq.\eqref{eq:MCObservable} with these sampled values, and bin the $N$ values into 50 evenly spaced bins that range from the minimum to the maximum gravitational wave frequencies allowed by the domain of the parameter space, giving the characteristic strain of the SGWB as provided in Eq.\eqref{eq:SGWBStrain}. The Monte-Carlo method is prone to highly fluctuating data due to the innate randomness, which is further exacerbated by the distributions for redshift and orbital frequency. We therefore repeat the simulation of all $N$ systems 25,000 times and calculate the average of $h_{\text{c,gwb}}^{2}(f)$. This significantly reduces sporadic fluctuations, especially at higher frequencies where there are less systems, however there still exists some noticeable fluctuations. These are merely artifacts of the realization of the Monte Carlo simulation; {\it they are not physical features}. It is important to note that specifically for the SE simulations, we do not account for the possible change in the observed gravitational wave frequency as a result of mass loss and orbital changes since the destruction timescale is much faster compared to the coalescence time.

\begin{figure}[htb]
    \centering
    \includegraphics[width=1.0\linewidth]{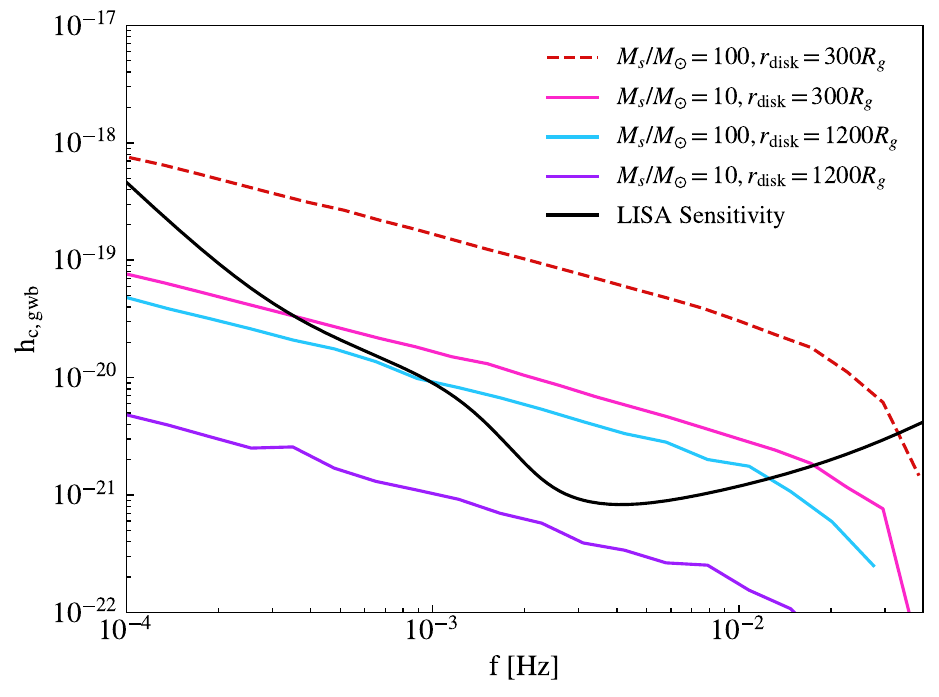}
    \caption{The inferred SGWB assuming all QPEs are BHEs. The red dashed line and pink line show the signal for EMRIs that form at $r_{\rm disk} = 300R_{\rm g}$ assuming all secondaries are 100 $M_{\odot}$ and 10 $M_{\odot}$, respectively. The fiducial rate used for these simulations are $R_{\rm BHE} = 6.07\times10^{-6}\text{ and }6.07\times10^{-7}$ gal$^{-1}$ yr$^{-1}$, respectively. The blue and purple lines show the signal for EMRIs that form at $r_{\rm disk} = 1200R_{\rm g}$ assuming all secondaries are 100 $M_{\odot}$ and 10 $M_{\odot}$, respectively. The fiducial rate used for these simulations are $R_{\rm BHE} = 2.37\times10^{-8}\text{ and }2.37\times10^{-9}$ gal$^{-1}$ yr$^{-1}$, respectively.} 
    \label{fig:BHE-SGWB}
\end{figure}

In Figure \eqref{fig:BHE-SGWB} we plot the SGWB for BHE-QPEs with $M_{s} = 100M_{\odot}, 10M_{\odot}$ at $r_{\rm disk} = 300R_{\rm g}, 1200R_{\rm g}$, along with the LISA sensitivity curve (presented in \cite{Robson2019}). The BHE-QPE systems have a significant contribution to the noise background in the range 1 mHz - 10 mHz, but also at frequencies as low as roughly 0.1 mHz. The results in Figure \ref{fig:BHE-SGWB} are consistent with the analysis performed by \cite{Chen2022} which finds that circular EMRIs have a dominant contribution to the SGWB over eccentric white dwarfs. This is likely due to a combination of factors, but likely the most prominent effect is that of the eccentricity spreading the signal into many different harmonics, expanding the signal to a broader range of frequencies and thereby lowering the overall strength of the signal in the LISA band. A noteworthy feature of Figure \ref{fig:BHE-SGWB} is the abnormally large signal generated for $M_{s} = 100M_{\odot}$ and $r_{\rm disk} = 300R_{\rm g}$, as it reaches nearly two orders of magnitude above the LISA sensitivity curve.

At first glance, one may interpret this as an indication that LISA will be overwhelmed by the EMRI background. However, both choices for $M_{s}$ and $r_{\rm disk}$ are highly conservative. As discussed previously, the semi-major axis of a BHE-QPE can be much larger than just 300 gravitational radii. A larger fiducial semi-major axis shifts more systems to lower frequencies, as well as reducing the total number of systems by lowering the inferred EMRI rate $R_{\rm BHE}$. These effects combined reduce the strength of the SGWB in the LISA band, which is evident when comparing to the results obtained by assuming $r_{\rm disk} = 1200R_{\rm g}$. Furthermore, the abnormally large signal is also a byproduct of the assumption that all BHE-QPEs have $M_{s} = 100M_{\odot}$. Larger secondary masses generate stronger gravitational waves, but such uniformity at this mass scale is certainly unphysical. Without knowledge of the secondary mass distribution however, we can only show the extremes. In this sense, the signal with $M_{s} = 100M_{\odot}$ and $r_{\rm disk} = 300R_{\rm g}$ is best understood as a rough upper-bound of the possible contribution to the SGWB by BHEs as opposed to an expected result. More physically reasonable results are the solid pink and blue lines that correspond to less conservative configurations.\\
\begin{figure}[htb]
    \centering
    \includegraphics[width=1.0\linewidth]{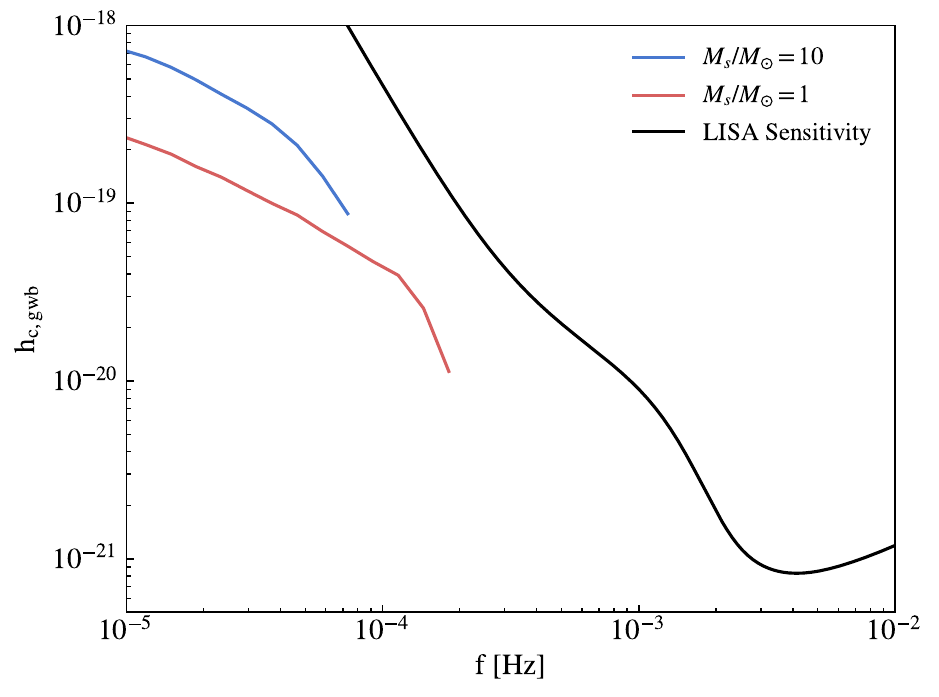}
    \caption{The inferred SGWB assuming all QPEs are SEs. The blue line shows the SGWB assuming every secondary is 10 $M_{\odot}$, the red line shows the SGWB assuming every secondary is 1 $M_{\odot}$, and the green line shows the SGWB assuming a log-uniform distribution for the secondary's mass. All simulations use $R_{\text{SE}} = 2.88\times10^{-6}$ gal$^{-1}$ yr$^{-1}$. Furthermore, we use $R_{\rm star} = R_{\odot}(M_{s}/M_{\odot})^{0.6}$ for the mass-radius relationship of the secondary, used in calculating the tidal radius.}
    \label{fig:SE-SGWB}
\end{figure}
\indent Figure \ref{fig:SE-SGWB} shows the possible contribution to the SGWB by SEs following the same plotting format as Figure \ref{fig:BHE-SGWB}. The contribution to the background is not detectable above the LISA sensitivity curve. A notable but expected feature of Figure \ref{fig:SE-SGWB} is the abrupt termination of the SGBW at sub-mHz frequencies. A main-sequence star will become tidally disrupted before it can reach an orbit that produces mHz gravitational wave frequencies and so will not contribute to the 1 mHz - 10 mHz band expected to host most resolvable EMRI signals. Furthermore, the dependence on $r_{\text{disk}}$ is not as dramatic as for BHE-QPEs since $R_{\rm SE}$ is independent of $r_{\text{disk}}$. Instead, increasing $r_{\text{disk}}$ lowers the theoretical minimum orbital frequency and consequently shifts the signal to lower frequencies.\\
\section{\label{sec:level4}Summary and Discussion}
In this paper we have presented a data-driven prediction of the SGWB using QPEs as tracers for circular EMRIs. In Section \ref{sec:level2} we derive simple and robust expressions to infer the EMRI formation rate from QPE observations. By assuming all QPEs and BHEs, then all QPEs are SEs, we obtain formation rates for BHEs and SEs, respectively. By making choices of fiducial numerical parameter values motivated by current QPE literature, we obtain fiducial rates $R_{\text{BHE}}$ and $R_{\text{SE}}$ that agree with rates predicted through different analyses, and these results in principle completely characterize the formation rates of the low-eccentricity QPE population. Section \ref{sec:level3} builds from the results of Section \ref{sec:level2}, using the Monte-Carlo method to simulate a mock population of EMRIs in order to infer their contribution to the SGWB. While both BHEs and SEs contribute to the sub-mHz background, only BHEs contribute to the 1 mHz - 10 mHz interval and so would impact direct observations of merging EMRIs by LISA. 

The background for the $M_{s} = 100M_{\odot}$ and $r_{\rm disk} = 300R_{\rm g}$ configuration appears to be unexpectedly large. As previously explained, this specific configuration of our model is the combination of two highly conservative estimates and is best interpreted as a theoretical upper bound; the remaining signals in Figure \ref{fig:BHE-SGWB} represent more physically reasonable results. From a different perspective, recent studies \cite{Witzany2026,Lui2026} reported that currently active QPEs have signal-to-noise ratios (SNRs) of roughly unity. The current active QPEs exist at observed gravitational wave frequencies below $10^{-4}$ Hz \cite{Witzany2026,Franchini2023,Zhou:2024vwj,Zhou2024,Zhou2024b,Giustini2024b}. As these systems are in the early inspiral phase, they are not expected to have high SNRs. As they finally evolve to higher frequency in the LISA sensitivity band, the expected SNRs could be orders of magnitude higher. In fact the current catalog of discovered QPEs, being not much larger than 10, is only a small fraction of EMRIs in the universe as only a small number of EMRIs can produce detectable QPEs. There are many more unresolvable systems that produce either weaker or stronger gravitational waves. Even if each system only produces gravitational waves of a strength similar to that shown in \cite{Witzany2026,Lui2026}, the combination of a large number of EMRIs can produce a detectable background in the LISA band.

While we were able to obtain order-of-magnitude results, a number of uncertainties in QPE data make rigorous constraints difficult to obtain, particularly when considering BHEs. The true extent of $r_{\rm disk}$ is not yet well understood. We choose to model it as a universal scale for simplicity, but its exact value is not clear with current QPE data. One should more broadly recognize that a single $r_{\rm disk}$ for all QPEs, despite its simplicity, may not be the most accurate prescription. In addition to the uncertainty surrounding $r_{\rm disk}$ is that for the secondary mass distribution which has not yet been deeply explored and contribute uncertainty to both $t_{\rm coal}(r_{\rm disk})$ and the orbital constraint $\lambda$ (and by extension the SGWB). Luckily, significant headway on these issues can be made by increasing the number of QPE detections. Analysis of many more QPE light curves and parameter inference will allow us to confirm whether these assumptions are representative of the QPE population, and this paper's model can be easily adjusted to account for new understandings of QPE structure and morphology.\\
\indent The results of this paper also set the framework for further advancements, both in work related to QPEs and to other fields of astrophysics and cosmology. An interesting avenue of research is the constraint of EMRI formation rates and formation channels. The model presented in Section \ref{sec:level2} does not assume \textit{apriori} a specific formation channel for low-eccentricity EMRIs, though the wet channel \cite{Pan2021a,Pan2021b,Sun:2025lbr,Zeng:2026ydj} appears to be the dominant formation channel of low-eccentricity EMRIs. An example of studying other formation channels is \cite{Mancieri2025,Allievi2026}, that consider two-body relaxation and GW dissipation in the dry channel as the mechanisms that drive initially eccentric secondaries into low-eccentricity orbits. 

\acknowledgements
X. C. acknowledges support from the National Natural Science Foundation of China under Grant No. 12473037.

\bibliography{bibliography}
\end{document}